\documentclass[]{article}
\usepackage{times,graphicx,a4wide, natbib}

\pdfoutput=1

\newcommand{\kms}{\rm{km \,s^{-1}}}
\newcommand{\msol}{M_{\rm \odot}}
\newcommand{\rsol}{R_{\rm \odot}}

\begin{document}
	
\title{Exoplanet Transits as the Foundation of an Interstellar Communications Network}
\author{Duncan H. Forgan$^{1,2}$}
\maketitle

\noindent $^1$SUPA, School of Physics and Astronomy, University of St Andrews \\
$^2$St Andrews Centre for Exoplanet Science \\

\noindent \textbf{Word Count: 5,366} \\

\noindent \textbf{Direct Correspondence to:} \\
D.H. Forgan \\
\textbf{Email:} dhf3@st-andrews.ac.uk \\

\newpage

\begin{abstract}

\noindent Two fundamental problems for extraterrestrial intelligences (ETIs) attempting to establish interstellar communication are timing and energy consumption.  Humanity's study of exoplanets via their transit across the host star highlights a means of solving both problems.  An ETI 'A' can communicate with ETI 'B' if B is observing transiting planets in A's star system, either by building structures to produce artificial transits observable by B, or by emitting signals at B during transit, at significantly lower energy consumption than typical electromagnetic transmission schemes.

This can produce a network of interconnected civilisations, establishing contact via observing each other's transits.  Assuming that civilisations reside in a Galactic Habitable Zone (GHZ), I conduct Monte Carlo Realisation simulations of the establishment and growth of this network, and analyse its properties in the context of graph theory.

I find that at any instant, only a few civilisations are correctly aligned to communicate via transits.  However, we should expect the true network to be cumulative, where a ``handshake'' connection at any time guarantees connection in the future via e.g. electromagnetic signals.  In all our simulations, the cumulative network connects all civilisations together in a complete network.  If civilisations share knowledge of their network connections, the network can be fully complete on timescales of order a hundred thousand years.  Once established, this network can connect any two civilisations either directly, or via intermediate civilisations, with a path much less than the dimensions of the GHZ.

\end{abstract}


\newpage

\section{Introduction}\label{sec:intro}


\noindent To date, the history of the search for extraterrestrial intelligence (SETI) has primarily been the history of radio SETI.  Since Frank Drake began the endeavour with Project Ozma in 1960, the vast majority of attempts to listen to or intercept signal transmissions from alien civilisations has focused on radio wavelengths (with a sizeable number of surveys also focusing on optical searches for extraterrestrial laser pulses).

\citet{TarterPlanetsLife2007} outlines the multi dimensional parameter space that both optical and radio SETI searches must investigate:

\begin{enumerate}
\item Distance
\item Direction
\item Signal Strength
\item Time
\item Polarisation
\item Frequency
\item Modulation/Information Content
\end{enumerate}

\noindent Naturally, some of these dimensions are far harder to investigate than others.  Signal timing is particularly problematic - SETI searches must be surveying the correct target as the transmission of the signal arrives at Earth.  Given that the most energy efficient method of transmission is in the form of highly collimated pulses \citep{Benford2010a,Benford2010}, SETI searches must target their instruments at the transmitter in the appropriate time interval when a pulse is arriving at Earth.  A signal beacon may choose only to emit a short duration pulse, with time delays of many years between pulses.  


This timing problem can permit a large number of transmitting civilisations, broadcasting with signal strength within the reach of current radio surveys, to be undetectable during the entire epoch of human SETI (which itself has only been fully active for a fraction of the last 60 years).

When we consider these issues, alongside the other aspects of the parameter space that SETI has only partially explored, it seems clear that ``classic'' SETI can only be successful if

\begin{enumerate}
\item a sufficiently large number of transmitting civilisations are present in the solar neighbourhood,
\item SETI searches can survey the local volume with high sensitivity
\item These surveys cover a wide range of signal polarisations and frequencies
\item SETI searches have relatively high survey cadence, combined with a long survey duration (of order a century or larger)
\end{enumerate}

\noindent Given the current political climate surrounding SETI, it seems unlikely that these criteria will be met unless a serendipitous detection is made (although modern privately funded surveys like Breakthrough Listen represent a significant step change in our efforts, see \citealt{Isaacson2017}).  Further to this, an energy efficient signal is likely to be highly collimated, which effectively guarantees our failure to intercept transmissions between two civilisations \citep{Forgan2014c}.  If we are to receive a radio transmission from ETI, it is likely that they have detected our presence and deliberately instigated contact.

And precisely how might ETI detect our presence? We can look to our own attempts to detect life beyond the solar system.  This is done principally through the science of extrasolar planet (exoplanet) detection.  There are several detection methods currently available to humanity, which to date have confirmed approximately 3000 exoplanets\footnote{http://exoplanets.org as of 20/06/17}.

The radial velocity technique monitors a star's spectrum for shifts in its spectral lines due to its reflex motion along the line of sight.  If a planet is present, the reflex motion will be periodic, as the star orbits the system's centre of mass \citep[see e.g.][]{Lovis2011}.  Gravitational microlensing efforts study the magnification of background stars by the gravitational field of an intervening lens star.  If the lens star hosts a planetary system, then the magnification is affected by the planet's own gravitational field \citep{Gaudi2011}.  

Direct imaging techniques use a variety of methods to remove the stellar flux from an image, and is able to detect photons emitted or scattered by the planet itself.  Currently, this technique requires the star and planet to be sufficiently separated on the sky for the stellar screening to be effective \citep{Traub2010}.

Most famously for SETI, exoplanets may be detected as they transit their host star.  Observers on Earth measuring the flux received by the star witness a characteristic ``dip'' as the planet passes between the observer and the star.  The depth of this dip indicates the stellar area that is obscured by the planet, which in turn indicates the planetary radius (as a function of stellar properties).  Observing transits at multiple wavelengths yields a planetary radius that varies according to how the planet's atmosphere absorbs incoming starlight, yielding crucial information about the planet's atmospheric composition and thermodynamic state \citep[see e.g][for a review]{Winn2011}.


It was quickly realised that exoplanet transits constitute a predictable, relatively strong, periodic, unpolarised electromagnetic signal over a wide range of frequencies.  In SETI terms, a great deal of the aforementioned multidimensional parameter space is collapsed.  If a transit signal contains evidence of intelligence, discovery of this intelligence is far more likely than a transmission directed at Earth without prior warning or coordination.

This has entered the public consciousness dramatically with the discovery of KIC 8462852, Boyajian's Star \citep{Boyajian2016}, which has a highly anomalous transit curve.  While many natural explanations have been applied to Boyajian's Star, such as large swarms of exocomets \citep{Bodman2016}, obscuration by the intervening interstellar medium \citep{Wright2016} or the recent consumption of an exoplanet \citep{Metzger2016}, an admittedly low probability explanation invokes the presence of alien megastructures obscuring the star \citep{Forgan_shkadov,Wright2015}.

The ``alien megastructure'' hypothesis is widely disbelieved by the SETI community, but it has sharpened the community's thinking on ways that SETI can operate beyond surveys for standalone electromagnetic signals.  Several authors have suggested means by which transits could provide a ``carrier'' of sorts for interstellar communications.  \citet{Arnold2005} proposed the construction of large geometric sheets (such as triangles) in orbit of a star, to produce transit curves distinguishable from those of exoplanets.  Appropriate positioning of these artificial structures would allow the transit curve to encode information, which could potentially last far beyond the lifetime of the transmitting civilisation \citep{Arnold2013}.

More recently, \citet{Kipping2016} proposed using laser pulses to modify the transit signal produced by the Earth.  If the ETI observing the transit has a known location, a modest laser pulse can be aimed at the observer to (for example) ``fill in'' the absorption lines added to the planet's transit transmission spectrum by biomarkers such as $\mathrm{O_2}$.  Conversely, the transit event can be used to time a deliberate laser pulse at the transit observer, in a sense ``piggybacking'' on the transit signal.

However an ETI may choose to do so, the transit can provide the means by which ETI in separate star systems synchronise efforts to initiate contact.  ETI ``A'' can establish contact with ETI ``B'' by adding signals to the transit curve induced by A (from the perspective of B).  As B can determine the epoch of transit for A, B directs its observations towards A at the appropriate time, and observes at the correct frequencies to receive any message. 

It is also reasonably straightforward for A to find targets for their transmission by checking a limited region of the sky for which, if an observer is located within this region, the transit of A can be measured.  This region has been calculated for Earth - the so-called Transit Zone, a band of sky centred on the Earth's ecliptic around half a degree wide.  Stars within this zone observing the Sun are appropriately aligned to see Earth transit \citep{Filippova1988,ConnHenry2008,Heller2016b}. 

Once contact is made, A and B now possess sufficient information to continue their communication either via transits, or via ``conventional'' electromagnetic transmissions.


This pairwise connection of ETI can be reproduced amongst any civilisation pair, provided either civilisation's host planet transits from the other's perspective.  As the stars hosting civilisations move around the galaxy, stars can enter each other's transit zones and their ETIs can establish connections.  

At this present instant, there are some 82 G/K stars within around 1 kpc in the \emph{Hipparcos} catalogue that reside in Earth's transit zone, i.e. these stars are positioned so that they can observe Earth's transit of the Sun.   Further analytic modelling suggests the complete catalogue could be nearly 3 orders of magnitude larger \citep{Heller2016b}.  The Sun has orbited the Galactic centre some 20 times since its formation (assuming a circular orbit at fixed galactocentric distance of 8 kpc and an orbital velocity of 220 $\kms$).  This implies that the Earth's transit zone has encompassed all stars with compatible orbits some 20 times during its existence, and hence the total number of stars that have entered the Earth's transit zone during the Earth's existence will number in the millions (if not more).

Over time, transit observations can allow the establishment of a network of connected ETIs.  In this work, I ask: \emph{what are the properties of this network? Is it robust? How effectively can messages delivered via transit propagate through this network, and should we expect it to replace more conventional methods?}


I display the results of Monte Carlo Realisation (MCR) simulations of alien civilisations communicating via transits in a Galactic Habitable Zone \citep[GHZ,][]{GHZ}, an annular section of the galactic disc that is sufficiently metal rich to permit habitable planet formation without suffering from overly hazardous local star formation.  I analyse the resulting network using the fundamentals of graph theory, and consider its overall efficacy as a means of transmission.  In section \ref{sec:method} I describe the MCR simulations, and the analysis performed on the resulting network; in section \ref{sec:results} I show the results for two different definitions of the GHZ, and in section \ref{sec:conclusions} I conclude the work.

\section{Method}\label{sec:method}

\subsection{Monte Carlo Realisation Simulations of ETI Communications via Transit}

\noindent We begin by generating a Galactic Habitable Zone (GHZ) of $N_*$ stars with intelligent civilisations, where we consider two forms of the GHZ: in the first, we follow \citet{GHZ} and consider an annular GHZ with inner and outer radii of 7 and 9 kiloparsecs (kpc).  In the second, we follow \citet{Gowanlock2011} and consider a GHZ as an annulus with inner and outer radii of 6 and 10 kpc respectively.  

Each star is assumed to be precisely solar (i.e. $M_*=\msol$, $R_*=\rsol$).  This means that transit probabilities for a given planetary system configuration are the same for all stars.  Larger stars have an increased transit probability

\begin{equation}
P_{\rm transit} \sim \frac{R_*}{a_p},
\end{equation}

\noindent but a reduced main sequence lifetime.  For this work, we follow the argument of \citet{Haqq-Misra2017} that suggests G stars may be the best sites for communicating civilisations.
 
Each star is randomly assigned a set of orbital elements, which remain fixed.  The semimajor axes of the stars $a_i$ around the galactic centre are exponentially distributed to simulate the Milky Way's surface density profile:

\begin{equation} 
P(a_i) \propto e^{-\frac{a_i}{r_S}},
\end{equation}

with the scale radius $r_S=3.5$ kpc\citep{Ostlie_and_Caroll}. The eccentricity distribution is uniform, under the constraint that a star's closest approach to the Galactic Centre must not be smaller than the inner radius of the GHZ (yielding maximum eccentricities of 0.22 and 0.4 for the Lineweaver and Gowanlock models respectively).  We also restrict the inclination of the orbits so that they do not exceed 0.5 radians.  The longitude of the ascending node (and the argument of periapsis) are uniformly sampled in the range $[0,2\pi]$ radians.  This naturally results in a Keplerian rotation curve, which is steeper than the roughly flat rotation curve expected for the Milky Way.  Given that we are only considering an annulus at most 4 kpc in extent, the difference in shear patterns between Keplerian and flat rotation curves can be safely neglected without changing the qualitative outcomes of our analysis.  Each star contains one and only one intelligent civilisation capable of communication.

The simulation is then run with a fixed timestep $\Delta t$, and the stars move in their fixed orbits.  We keep the ratio $t_{\rm max}/\Delta t=1000$, i.e. if the simulation duration is 1 Gyr, then $\Delta t = 1$ Myr, and if the duration is 1 Myr, then $\Delta t = 1$ thousand years. Each star has a planetary system assigned to it, with a randomly assigned inclination relative to the Galactic plane.  As stars move relative to each other, the ability for civilisation pairs to initiate contact via transits will appear and disappear.

For simplicity, planets are uniformly assigned circular orbits, with a semimajor axis around their host star between 0.1 and 100 AU.  Tests run where the semimajor axis is sampled from the currently observed exoplanet semimajor axis distribution show little difference in results.  We assume that the orbital period of the planet (between weeks to centuries depending on the semimajor axis) is much shorter than the time that external observers survey the system, and that the transits remain visible to said observers for much longer timescales than the orbital period.

\begin{figure}
\begin{center}
\includegraphics[scale=0.4]{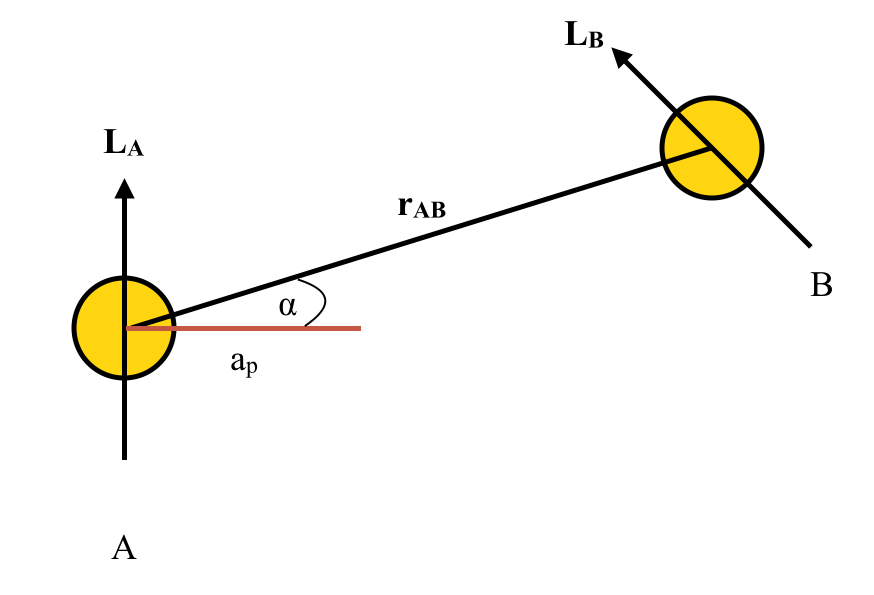}
\end{center}
\caption{The geometry of detecting transits.  Star B can detect transits from star A if the angle $\alpha$ is sufficiently small (see text). \label{fig:transit}}
\end{figure}

\noindent At any given instant, we can determine if star A is in star B's transit zone by the following (Figure \ref{fig:transit}).  Firstly, we calculate the separation vector $\mathbf{r}_{AB} = \mathbf{r}_{A} - \mathbf{r}_{B}$.  We then compute the angle $\alpha$ between the separation vector and the orbital plane of the planetary system A, defined to be perpendicular to the angular momentum vector, $\mathbf{L}_{A}$.  For a transit to be visible, the projected distance in the limit of small $\alpha$:

\begin{equation}
R = \alpha a_{p},
\end{equation}

\noindent must be smaller than the critical distance

\begin{equation}
R_c = R_* + R_p,
\end{equation}

\noindent where we fix $R_p=1 R_{\oplus}$.  This is a relatively slack condition, as grazing transits, where the planet only barely covers the stellar disc, are regarded as detectable.  If $R<R_c$, we record the connection between the two systems at this timestep.

Every timestep produces an undirected graph identifying all the connections currently active.  At the end of the simulation we also record a cumulative graph, where every connection made between two stars over the course of the simulation is collected.

\subsection{Analysing the Transit Communication Network}

\subsubsection{Defining the Graph - Vertices, Edges and Components}

\noindent The communication network can be regarded as an undirected graph $G$, composed of $N_{\rm vertices}$ \emph{vertices} ($N_{\rm vertices} \leq N_*$), and $N_{\rm edge}$ \emph{edges}, where each edge represents a connection made between two stars via transits.  We define a \emph{path} $P$ as a set of edges that connect one vertex with another.

We analyse the following graph properties.  Each graph contains a number of subgraphs referred to as connected components.  A connected component consists of a subset of vertices that can be connected by an unbroken path, and the edges constituting said paths. The number of connected components $N_{comp}$ is a simple but effective measure of the connectivity of the network.  If $N_{comp}=1$, then all vertices are connected to each other through a path.  A larger value of $N_{\rm comp}$ indicates a collection of networks, rather than a single network (see e.g. the left panel of Figure \ref{fig:graphs}, which has $N_{\rm comp}=2$).

We will refer to the number of vertices that have established a connection as $N_{\rm members}$, and the number of vertices that have established no connections at all as $N_{\rm isolated}$, with $N_{\rm members} +N_{\rm isolated} = N_*$.

\subsubsection{The Minimum Spanning Tree/Forest}

\noindent If we have a graph $G$ that is fully connected ($N_{\rm comp}=1$), then the minimum spanning tree (MST) is a subset of $G$ that contains all the vertices, and a subset of the edges of $G$.  The edges are selected such that all vertices remain connected, and the total edge weight is minimised.   We define the edge weight as the distance between vertices connected by the edge.  

\begin{figure*}
\begin{center}$
\begin{array}{cc}
\includegraphics[scale=0.35]{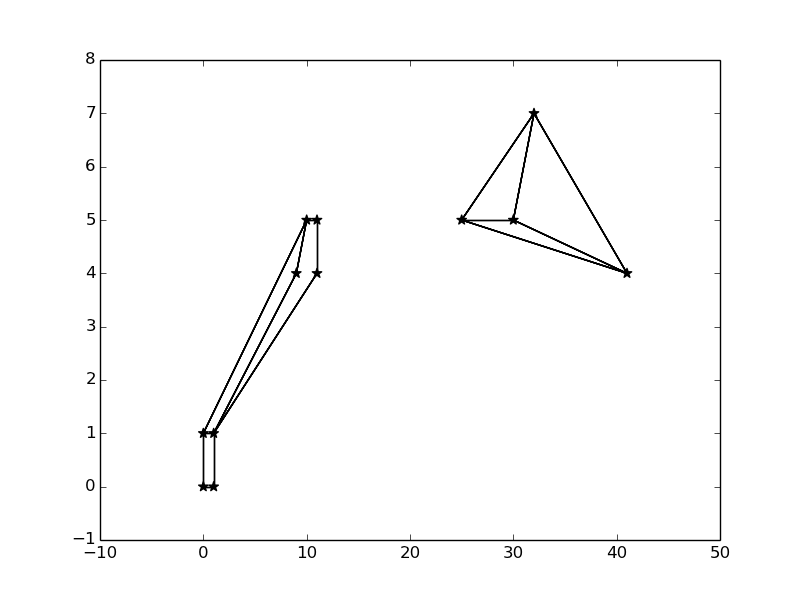} &
\includegraphics[scale=0.35]{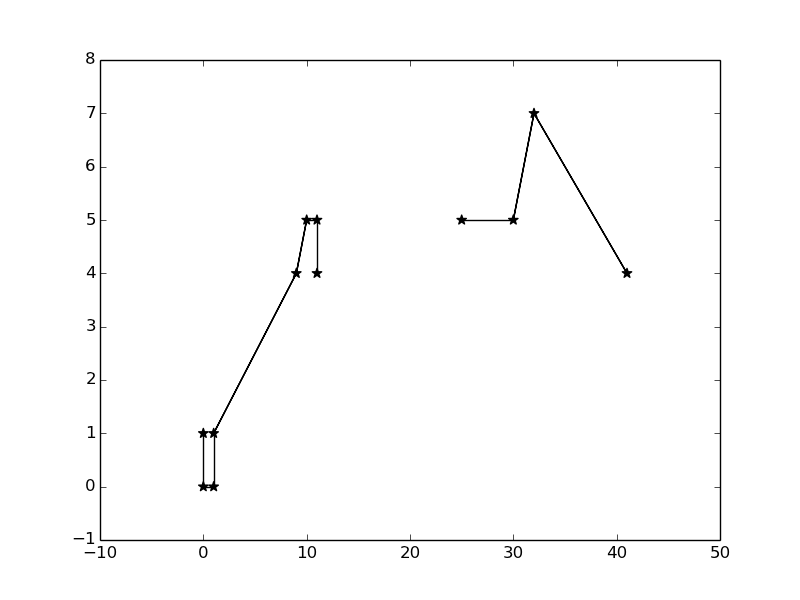}
\end{array}$
\end{center}
\caption{Left: An example of a graph with two connected components.  Right: The minimum spanning forest of this graph.  \label{fig:graphs}}
\end{figure*}

Broadly speaking, the minimum spanning tree of $G$ represents the minimum distance a signal would need to travel along a network to reach every member.  If $G$ is not fully connected ($N_{comp}>1$), we can compute a minimum spanning forest (MSF), which in essence is a set of minimum spanning trees, with one tree per connected component of $G$ (right panel of Figure \ref{fig:graphs}).  

We compute MSFs for our graphs using the DJP algorithm \citep{Jarnik1930,Prim1957,Dijkstra1959}.  This algorithm is designed for fully connected graphs - in our implementation, it is run on each connected component to deliver a MST for each, and in their combination the MSF.

\subsubsection{The Minimum Path Between Vertices}

If two vertices $a$ and $b$ exist within the same connected component, we can identify a minimum path $P_{\rm min}(a,b)$ between the two.  We do this using the $A^*$ algorithm \citep[see e.g.][]{Zeng2009}.  The algorithm attempts to find the minimum path by reducing the distance to the target vertex.  Briefly, the algorithm is described as follows: to find the next vertex on the path from vertex $i$, all vertices $\{k\}$ that share an edge with $i$ have the following weight calculated:

\begin{equation}
f(k) = g(k) + h(k),
\end{equation}

where $g(k)$ is the distance from the first vertex $a$ to vertex $k$, plus a heuristic $h(k)$.  Typically, $h(k)$ is the distance from $k$ to the target:

\begin{equation}
h(k) = \left|\mathbf{r}_k - \mathbf{r}_b\right|,
\end{equation}

and we make the same choice for $h(k)$ here.

%
%
%
%

\section{Results}\label{sec:results}

\noindent In the following, we consider the network in two guises.  Firstly, we consider the \emph{instantaneous} network, which indicates how many civilisations are able to communicate with each other via transits at a given moment in Galactic history.

Secondly, we consider the \emph{cumulative} network, which allows civilisations to maintain their connection to each other once they move out of alignment to communicate via transits.  

\subsection{Instantaneous Network Properties}

\begin{figure*}
\begin{center}$\begin{array}{cc}
\includegraphics[scale=0.35]{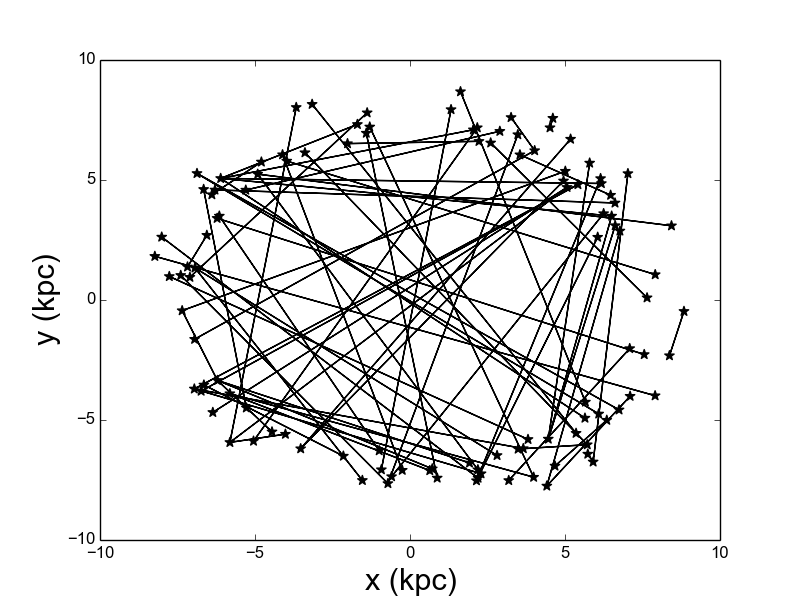}
\includegraphics[scale=0.35]{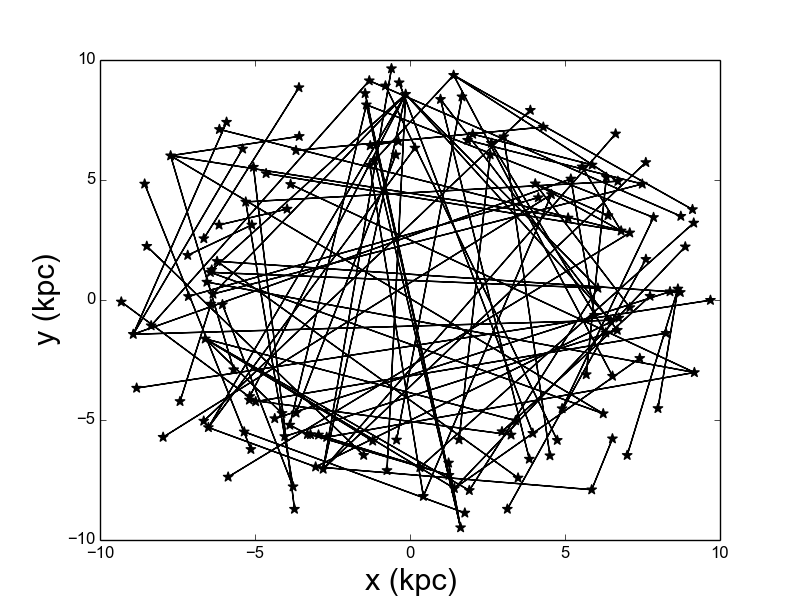}
\end{array}$
\end{center}
\caption{The minimum spanning forest (MSF) of the instantaneous transit communication network at $t=500$ Myr \label{fig:halfway}}
\end{figure*}

\noindent Figure \ref{fig:halfway} shows the state of the instantaneous transit communication network in the Lineweaver GHZ (left panel) and the Gowanlock GHZ (right panel), for a single Monte Carlo Realisation, at $t=500$ Myr (i.e. halfway through the run).  We plot the minimum spanning forest (MSF) of the networks for the sake of clarity.  As can be seen, the MSF demands multiple connections that span the Galactic Centre.  This is in part due to the large number of connected components - at this instant, $N_{comp} = 36$ for the Lineweaver GHZ, and $N_{comp}=39$ for the Gowanlock GHZ.  The networks also leave a relatively large number of unconnected civilisations at this particular instant (around 350 for both cases).

\begin{figure*}
\begin{center}$
\begin{array}{cc}
\includegraphics[scale=0.35]{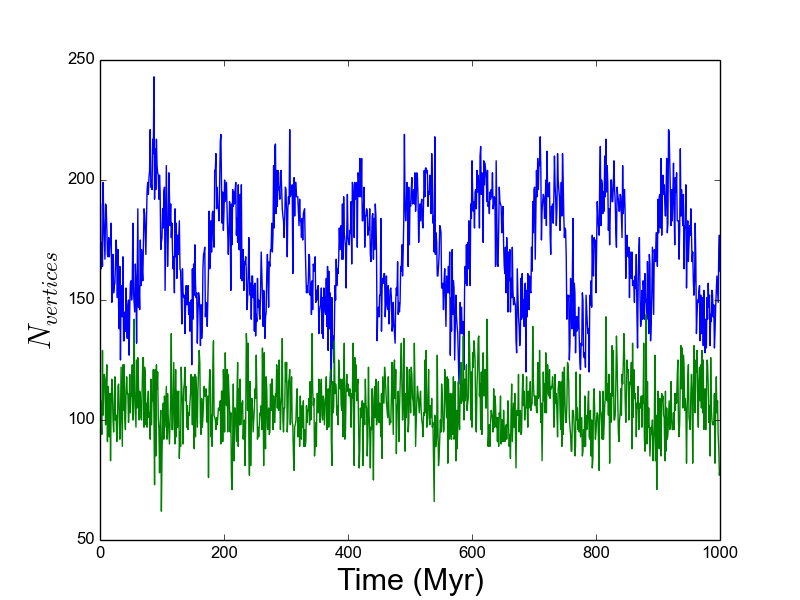} &
\includegraphics[scale=0.35]{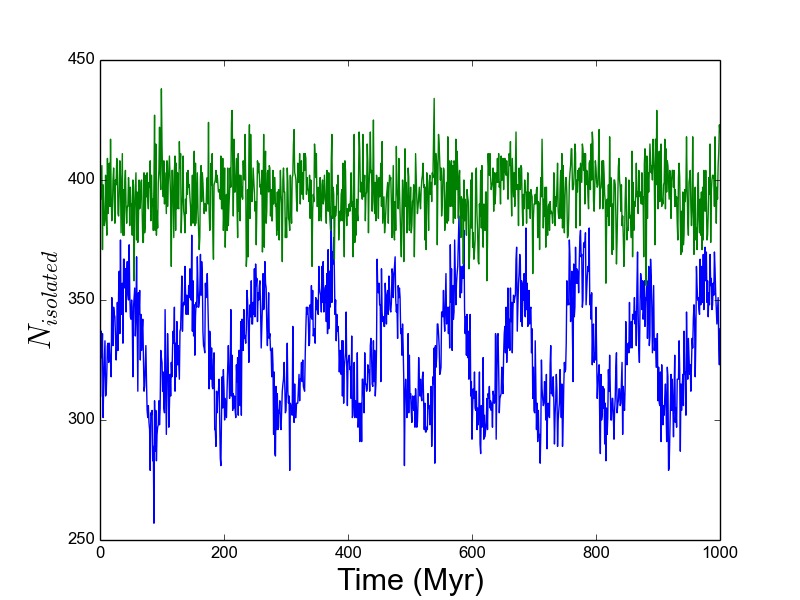} \\
\includegraphics[scale=0.35]{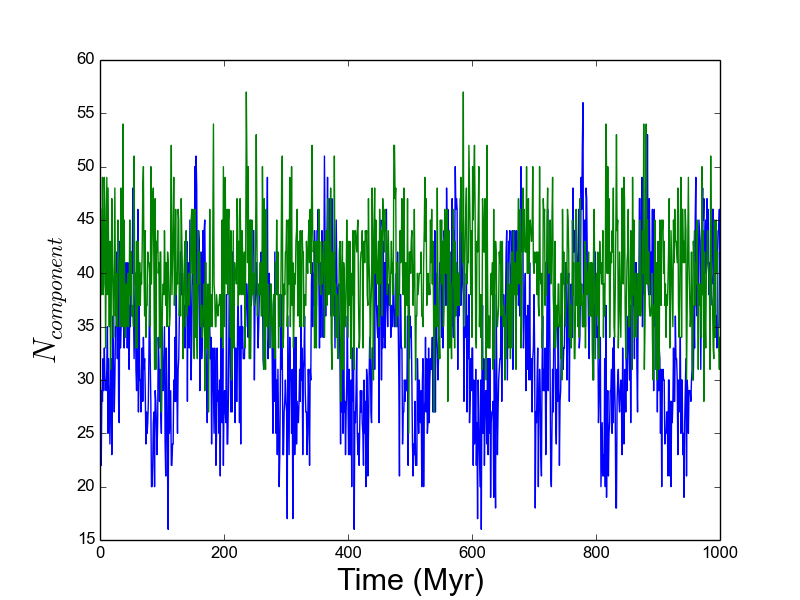} &
\includegraphics[scale=0.35]{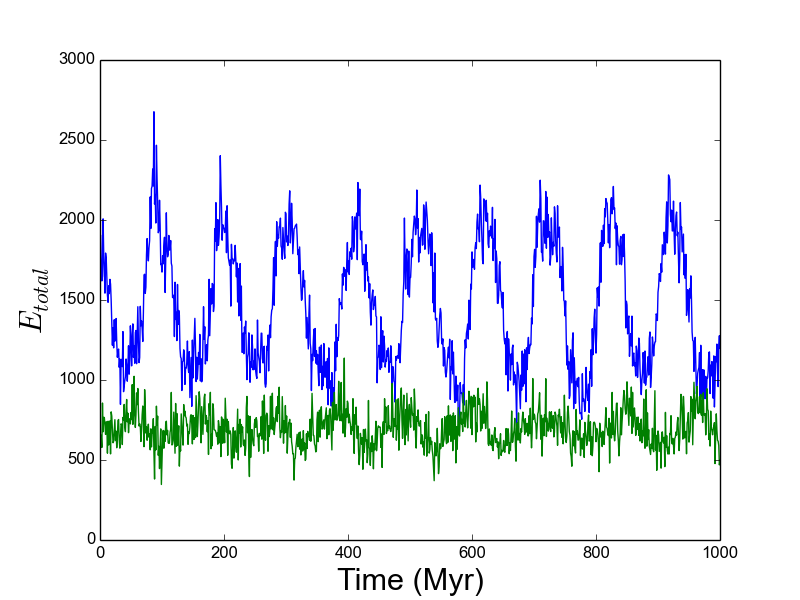}
\end{array}$
\end{center}
\caption{The instantaneous properties of the transit communication network for two different Monte Carlo Realisations, assuming the Lineweaver Galactic Habitable Zone.  Top left and right: the number of connected and isolated vertices, bottom left and right: the number of connected components, and the total length of edges in the network. \label{fig:inst_lineweaver}}
\end{figure*}

We can see this more clearly by investigating the evolution of the graphs' global properties with time.  Figure \ref{fig:inst_lineweaver} shows two different realisations in the Lineweaver GHZ, and it is immediately clear that the graph behaviour can vary quite widely between realisations.  Both show some periodic variations in the number of civilisations connected to another civilisation (top left plot), and equally the same variation in completely unconnected civilisations (top right).  This variation is extremely weak in one, and much stronger in the other.  The period of the oscillation is approximately 150 Myr, which roughly corresponds to the rotation period of the GHZ zone inner edge at 7 kpc \citep[see e.g.][]{Bhattacharjee2013}.

In both runs, the number of connected components is always in the tens, and typically between 35 and 40.  One run shows a relatively poor quality network, with the number of isolated civilisations at any time being of order 400, with the other showing a level of isolation nearer to 300.  However, we can see that the more connected network also has a proportionately larger total edge length.  If we divide the total edge length by the number of civilisations in the network, one run has values of 8-10 kpc per connected civilisation, as opposed to slightly lower values of 5-8 kpc in the other.  

\begin{figure*}
\begin{center}$
\begin{array}{cc}
\includegraphics[scale=0.35]{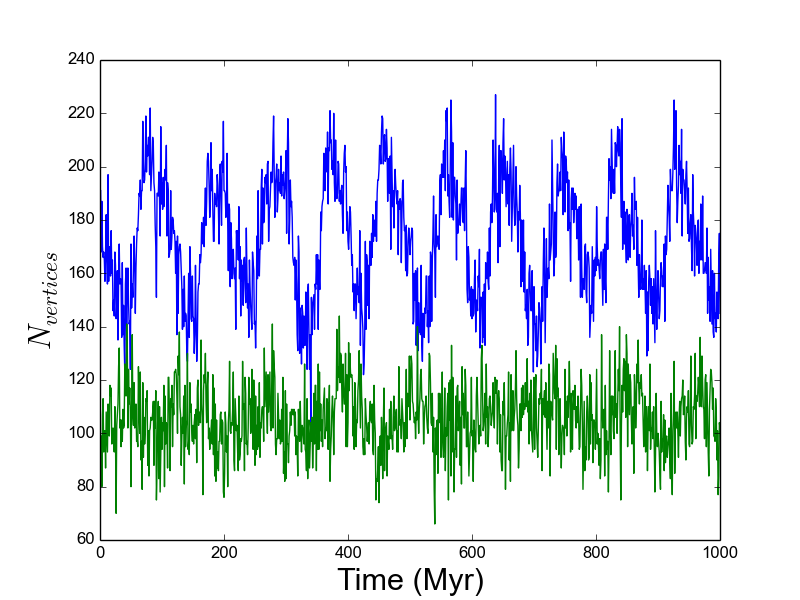} &
\includegraphics[scale=0.35]{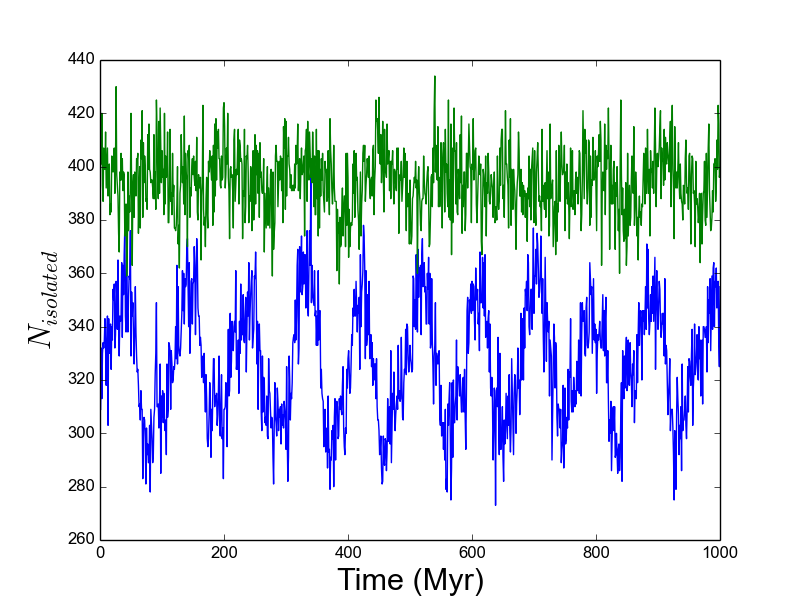} \\
\includegraphics[scale=0.35]{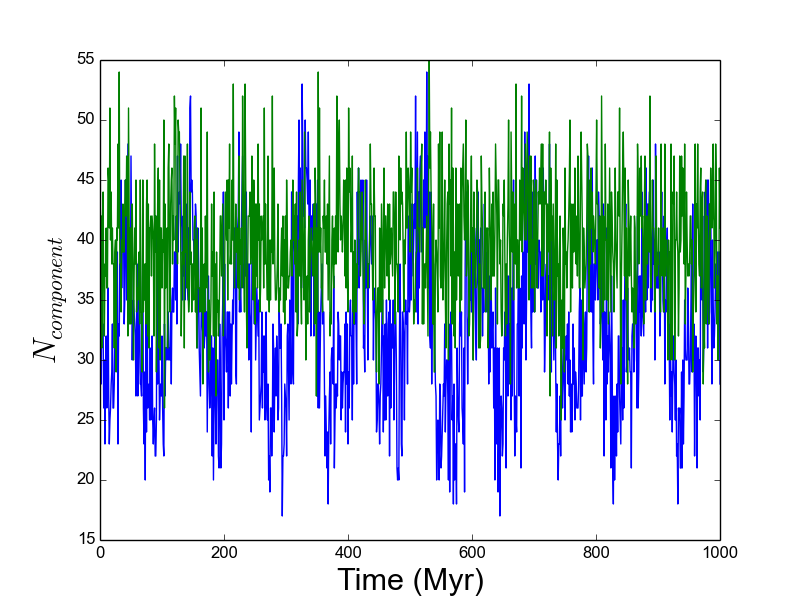} &
\includegraphics[scale=0.35]{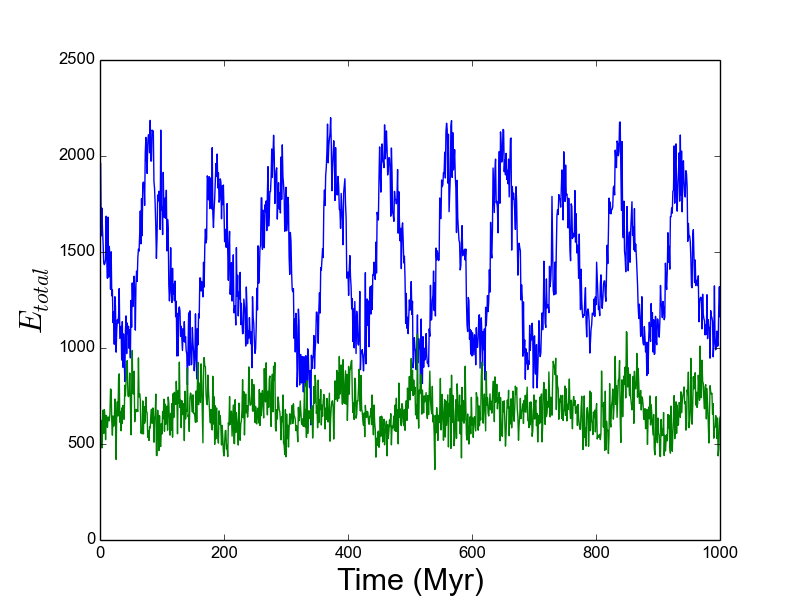}
\end{array}$
\end{center}
\caption{The instantaneous properties of the transit communication network for two different Monte Carlo Realisations, assuming the Gowanlock Galactic Habitable Zone.  Top left and right: the number of connected and isolated vertices, bottom left and right: the number of connected components, and the total length of edges in the network. \label{fig:inst_gowanlock}}
\end{figure*}

Similarly, Figure \ref{fig:inst_gowanlock} shows two realisations in the Gowanlock GHZ again that two different realisations of the simulation can produce quite different behaviour.  The periodicity seen previously remains, with a reduced period as the inner radius of the GHZ is also reduced. 

In short, the properties of the instantaneous communication network are relatively insensitive to the definition of the GHZ.  The number of civilisations connected at any one time is typically quite a low fraction of the total number.  Depending on good or bad fortune, the properties of the instantaneous network can evolve quite significantly on timescales linked to the Milky Way's rotation curve.

\subsection{Cumulative Network Properties}

\noindent Let us now consider the cumulative network - i.e., we now assume once a connection between two civilisations is forged, it remains regardless of the orientations of their planetary systems and positions of their host stars.  In the previous section, we learned that the properties of the instantaneous network change significantly from realisation to realisation.  We find that for the cumulative networks, regardless of the GHZ model, all realisations converge on a similar behaviour - a highly connected network, where every civilisation is eventually added to the network, and $N_{comp}=1$.

\begin{figure*}
\begin{center}$
\begin{array}{cc}
\includegraphics[scale=0.35]{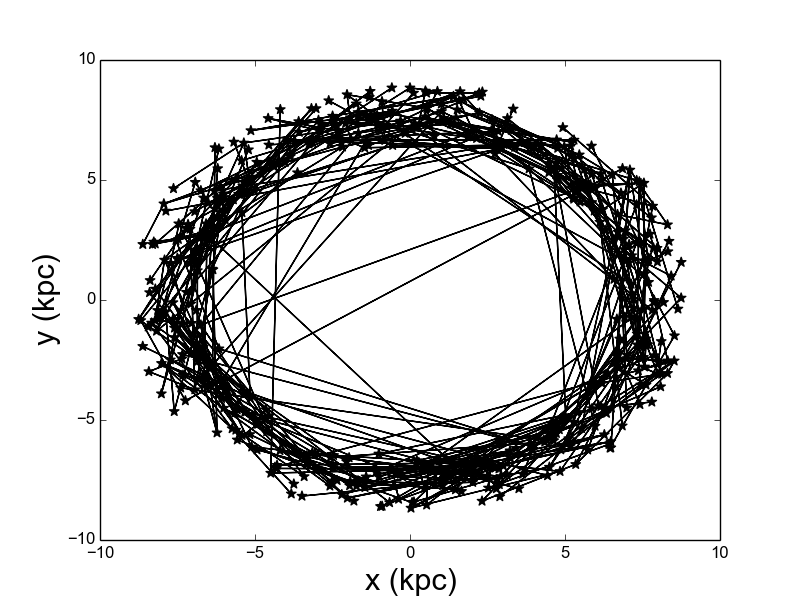} &
\includegraphics[scale=0.35]{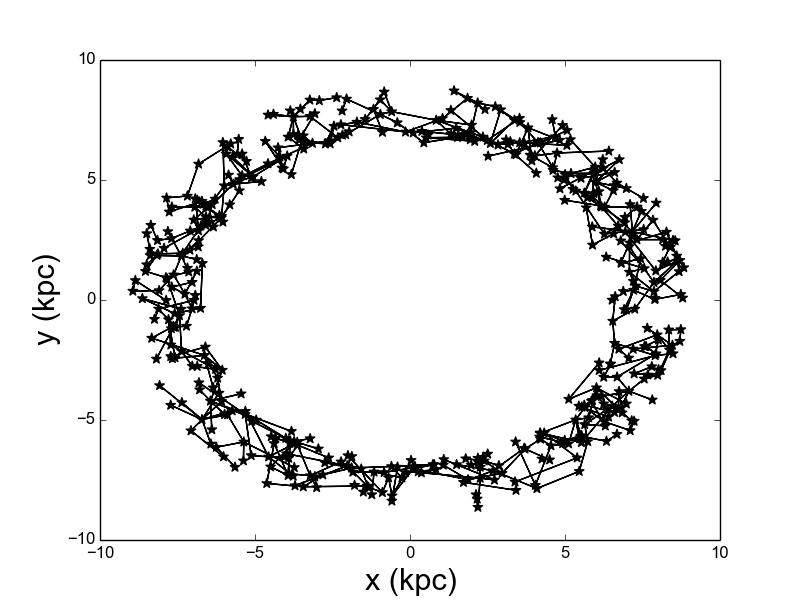} \\
\includegraphics[scale=0.35]{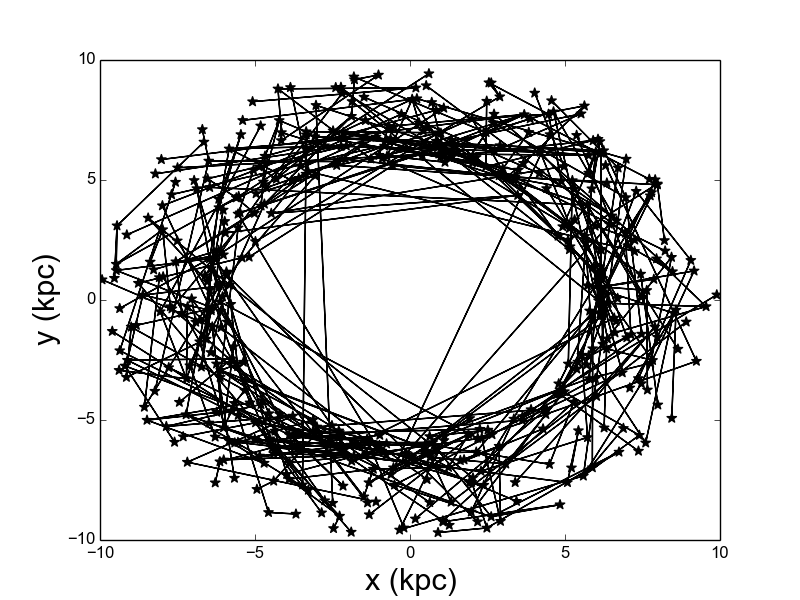} &
\includegraphics[scale=0.35]{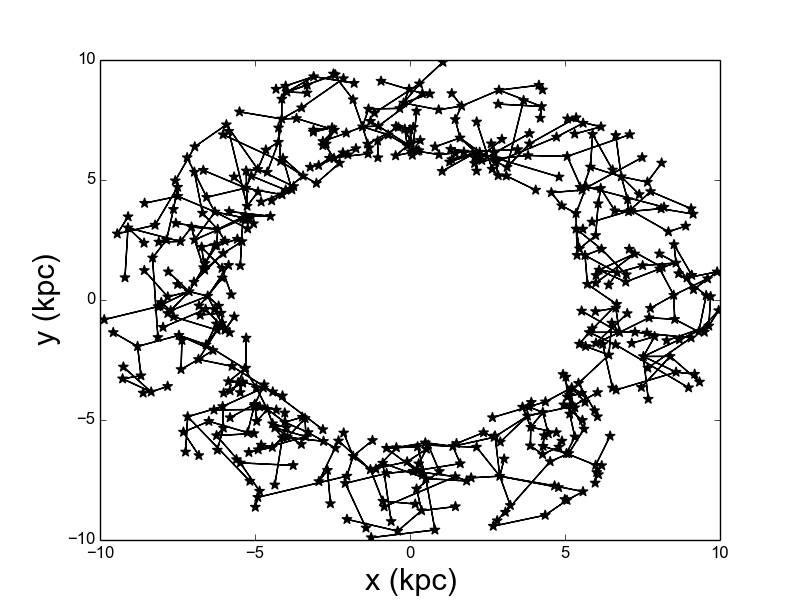}

\end{array}$
\end{center}
\caption{The minimum spanning forest (MSF) of the transit communication network after 1 Myr (left) and after 1 Gyr (right).    Runs using the Lineweaver GHZ occupy the top row, and the Gowanlock GHZ runs occupy the bottom row. \label{fig:cumulative_MSF}}
\end{figure*}

Figure \ref{fig:cumulative_MSF} shows the minimum spanning trees for the cumulative network of a single realisation, using the Lineweaver GHZ (top row) and the Gowanlock GHZ (bottom row).  We run with the same parameters with a maximum runtime of 1 Myr (left) and 1 Gyr (right).  In both cases, all civilisations are connected, and $N_{comp}=1$.  With 1 Myr of evolution, a sufficiently large enough number of connections have been established so that all civilisations are contained within the network, but many of these connections are still long range (spanning the Galactic Centre).  As we can see in the left panels of Figure \ref{fig:cumulative_MSF}, the MSF of each network still contains some of these long ranging connections.  

After 1 Gyr, nearly every civilisation has connected to each other at least once.  With this much larger set of possible connections, the minimum spanning forest no longer needs to traverse the Galactic centre, instead using the multitudinous smaller length connections to deliver a network with significantly shorter total connection distance.

\begin{figure*}
\begin{center}$\begin{array}{cc}
\includegraphics[scale=0.35]{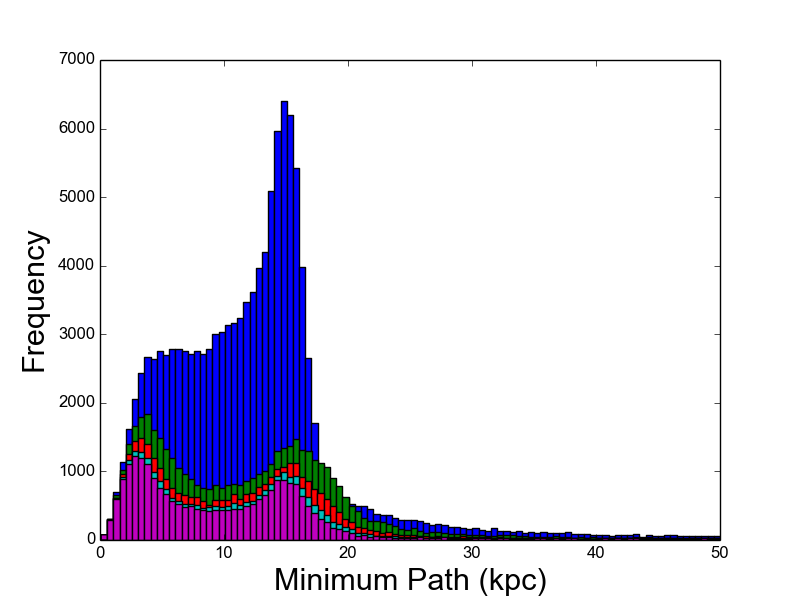} &
\includegraphics[scale=0.35]{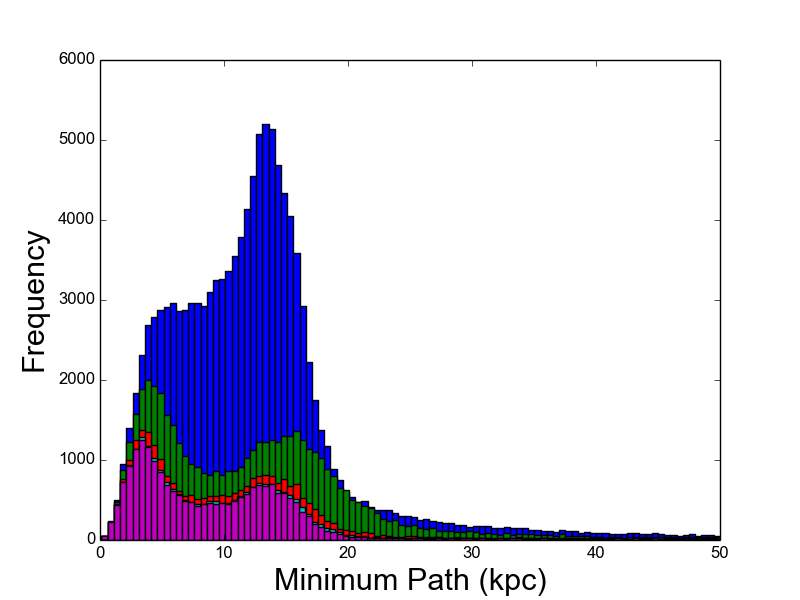}
\end{array}$
\end{center}
\caption{The distribution of minimum paths between vertices in the network for five realisations of the Lineweaver GHZ (left), and five realisations of the Gowanlock GHZ (right). \label{fig:minpaths}}
\end{figure*}

We can now ask: what is the typical minimum path $P$ along this network between any two civilisations? We calculate the minimum path between all pairs of civilisations in the network.  Figure \ref{fig:minpaths} shows the distribution of $P$ (in kpc) between all pairs for five realisations for both GHZ models.  In principle, the maximum civilisation separation is $2R_{\rm out}$ (18 and 20 kpc for the Lineweaver and Gowanlock models).  We can see the distribution of $P$ peaks around $2R_{\rm out}$ in both cases.  A second peak is visible around 3-5 kpc.

As an illustrative example, let us imagine that we wish to send a message along the network to a civilisation that also resides at a galactocentric distance of $R=8$ kpc, but on the opposite side of the Galactic Centre to Earth.  From Figure \ref{fig:cumulative_MSF}, we can see the minimum spanning tree is roughly concentric in shape.   We can therefore estimate the path length of the signal as simply $\pi R$, (i.e. half the circumference of a circle of radius $R$).  This gives a distance of approximately 25 kpc (and a travel time of approximately 0.08 Myr).  A direct transmission through the Galactic Centre only has a path length of 16 kpc (travel time $\sim$0.05 Myr), but is unlikely to reach its target due to the obscuring nature of the Galactic Centre.

\subsection{Sharing Address Books - Secondary Connections and Rapid Network Growth}

\noindent So far, we have also neglected secondary connections in this analysis.  Until now, when A connects with B, the ``address books'' of A and B (the vertices that A and B are connected to) are not shared.

We now consider the case where address books are shared, i.e. when A and B connect, A shares with B the location of other members of the network (as well as members A was connected to in the past).  In other words, whenever a vertex is added to the graph, it immediately connects to any vertices that were part of the network at any point in the past.

\begin{figure*}
\begin{center}$\begin{array}{cc}
\includegraphics[scale=0.35]{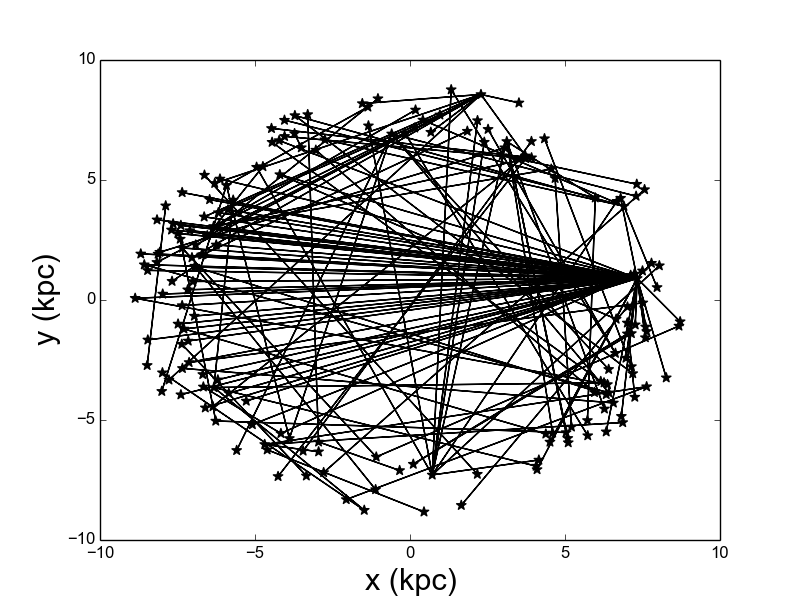} &
\includegraphics[scale=0.35]{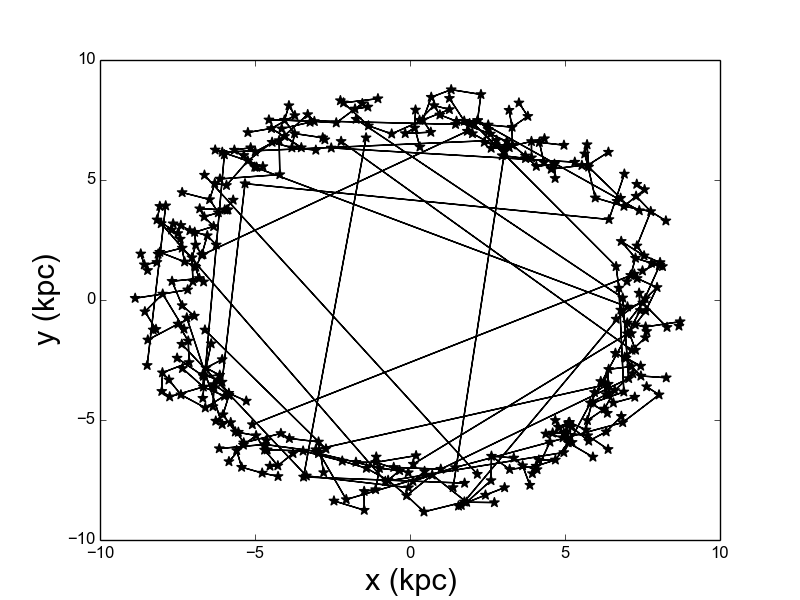}
\end{array}$
\end{center}
\caption{The effect of sharing members' ``address books''.  Left: the minimum spanning forest of a network without address sharing in the Lineweaver GHZ after 100 kyr, with 180 members (and 29 connected components).  Right: the same network with address sharing, with 374 members (25 components). \label{fig:sharing}}
\end{figure*}

This results in a dramatic increase in the growth rate of the cumulative network. We rerun the Lineweaver GHZ for 1 Myr, and find that after 100 kyr, the network already contains 374 members, and contains all members after 300 kyr.  We compare the non-sharing and sharing runs at 100 kyr in Figure \ref{fig:sharing}.

It is immediately plain that address book sharing is an effective strategy in building complete networks an order of magnitude faster than would be possible otherwise.

\section{Discussion}\label{sec:discussion}

\subsection{Limitations of the Analysis}

\noindent We have fixed the stellar orbits in this work, which is a clear oversimplification.  The orbits of stars will evolve according to the Galactic potential, resulting in radial mixing.  These dynamical effects not only move stars in and out of an annular GHZ \citep{Vukotic2016}, but are instrumental in altering the GHZ's morphology \citep{Forgan2017}.  We have also not considered the highly clustered nature of the Milky Way's stellar content.  Transit networks in globular clusters will be significantly denser than those in field stars, and the gravitational potential well of a cluster will alter stellar orbits relative to the Galactic Centre at an even greater rate.  As a result, we might expect a real transit network to be highly substructured, and bend and flex according to these gravitational forces.

These forces are also at play on planetary scales.  The orbital elements of planets experience perturbations from their neighbouring planets, from companion stars if they are present, and from close encounters with neighbouring stars (provided that the encounter is sufficiently close).  The visibility of transiting planets can therefore vary on timescales significantly shorter than those due to stellar motions, if the orbital inclination or ascending node is shifted.  For example, the circumbinary planet Kepler-16ABb will no longer transit either star from Earth's viewpoint as of 2018 \citep{Doyle2011}.

We should also note that we do not consider binary systems at all in our analysis.  A significant fraction of main sequence stars are found in binaries \citep{Duquennoy1991,Raghavan2010}.  The transit probabilities for circumbinary systems may be significantly more favourable \citep{Martin2015}, even if the rapid precession of periapsis means that transits are only visible for a short duration (in cosmic terms).

In this work, we have assumed that civilisations have a lifetime greater than 1 Gyr.  Even if the typical civilisation lifetime was around 1 Myr, it seems our networks would eventually incorporate every member, provided that most civilisations came into existence at a similar time (although as we have seen, the MSF of such a network would have a larger total edge length).

A civilisation's death may not prevent them from sending messages in the network - for example, in the case of \citet{Arnold2013}'s suggestion of large orbiting structures as communication, messages sent using this technique will continue as long as the orbit of the structure is stable.  A structure designed to be stable without propulsion may still be transmitting long after its makers are extinct.  Of course, if a message from another ETI is to be relayed through this node in the network, such an effort will fail.  The deep interconnectivity of our simulated networks after 1 Gyr suggests that a failure such as this might be easily circumvented by utilising a different route.  Future analysis of this network should pay careful attention to this issue to confirm the network's robustness in the face of civilisation destruction.

\subsection{Advantages and Disadvantages of using the Transit Network}

\noindent Some of the advantages of the transit network have already been described in this article.  Signals transmitted along the network are likely to require significantly less energy (after the initial construction outlay).  The \citet{Arnold2005} structure approach is energetically intensive when the structures are being built, but the messaging itself only requires thrust for station-keeping, sending a signal that can be received at significant distance without the power required to send an electromagnetic transmission.

Equally, a high level of technological sophistication is required to determine the transit of an exoplanet.  Civilisations must surpass a relatively high bar to join the transit network - the height of the bar is determined by the signal strength (i.e. the size of the orbiting megastructure, or the power of the laser signal).  Communicating via transits may allow civilisations to  ``screen out'' less developed neighbours from contact.

Once acquired, the transit network signals are extremely predictable, with each transmission corresponding to an orbital period.  Acquiring several signals would allow the receiver to establish this period, and plan their own transmissions accordingly.

We can estimate the number of two way exchanges ETIs can conduct using transits by comparing the message travel time and the timescale on which the ETIs leave each other's transit zone.  A message with a path of 20 kpc (the diameter of the GHZ) has a total travel time at lightspeed of just under 0.06 Myr.  If we assume a relatively short timescale on which both ETIs remain in the transit zone of 100,000 years (which is approaching the timescale on which both secular evolution of planetary orbits and the star's orbit become important), then a total of 30 exchanges can be made.  This of course does not forbid a continuing conversation by other means.


The periodicity of this signal is a double edged sword - signals can only be sent when the receiver begins to see the transit.  Megastructures can be moved to alter the epoch of transit relative to an observer, but this poses its own problems in terms of synchronising transmission and reception.

While transits require high precision photometry to observe, once this technology is achieved the signals can be received by any civilisation conducting an exoplanet transit survey (as humans are currently).  Eavesdropping on direct transmissions between two civilisations is hard unless the transmission is relatively uncollimated \citep{Forgan2014c}.  Transits are a relatively uncollimated signal (i.e. they are visible at a small angular separation from the line of sight) compared to say a laser beam.  Using transits as a means of communication may permit intended or unintended off-axis transmissions - in effect, a form of highly directed signal leakage.  This may present a more effective means by which humans can scout the local interstellar volume for civilisations transmitting to each other and avoiding the Earth \citep[see also][]{SKA,Forgan2016d}.

%

\section{Conclusions}\label{sec:conclusions}

\noindent We have shown that extraterrestrial intelligences (ETIs) can build a highly robust communication network, utilising the fact that ETIs can observe planets in other star systems using the transit method.  Civilisation `A' can communicate with civilisation `B' by modifying the transit signal from A's home planet, as observed by B.

Using graph theory, we have demonstrated that all civilisations in a Galactic Habitable Zone can establish a fully connected network within a million years, where all civilisations are connected to each other, either directly or via intermediate civilisations.    Given further time (or judicious information sharing) this network can establish a minimum spanning tree that has a very low total connection length.  This network would require far less energy to transmit data, and the range of any signal is effectively the entire extent of the volume occupied by ETIs, provided that the message is retransmitted by intermediate civilisations.

Given that these transit signals are likely to be polychromatic, highly periodic (and potentially information dense), exoplanet transit surveys become a direct and powerful SETI search tool.  As transit surveys encompass a growing number of stars with higher precision and cadence, we can place direct constraints on the membership and connections of any putative transit network in the Solar neighbourhood by merely continuing our search for extrasolar planets.


\section*{Acknowledgments}

The author gratefully acknowledges support from the ECOGAL project, grant agreement 291227, funded by the European Research Council under ERC-2011-ADG, and the STFC grant ST/J001422/1.  The author warmly thanks Ren{\'e} Heller for a careful and extremely helpful review of this manuscript.

\bibliographystyle{mn2e} 
\bibliography{transit_bridges}

\end{document}